# Can we Compute the Similarity Between Surfaces?

Helmut Alt    Maike Buchin [*]

October 30, 2018

**Abstract**

A suitable measure for the similarity of shapes represented by parameterized curves or surfaces is the Fréchet distance. Whereas efficient algorithms are known for computing the Fréchet distance of polygonal curves, the same problem for triangulated surfaces is NP-hard. Furthermore, it remained open whether it is computable at all.

Using a discrete approximation we show that it is *upper semi-computable*, i.e., there is a non-halting Turing machine which produces a monotone decreasing sequence of rationals converging to the Fréchet distance. It follows that the decision problem, whether the Fréchet distance of two given surfaces lies below a specified value, is recursively enumerable.

Furthermore, we show that a relaxed version of the Fréchet distance, the *weak Fréchet distance* can be computed in polynomial time. For this, we give a computable characterization of the weak Fréchet distance in a geometric data structure called the *free space diagram*.

## 1 Introduction

Suitable distance measures for comparing the similarity of shapes are an important issue in application areas such as computer vision and pattern recognition. The distance measure that comes to mind first is the Hausdorff-distance which, between two compact sets $A, B \subset \mathbb{R}^d$, is defined as

$$\delta_H(A,B) = \max \big( \max_{a \in A} \min_{b \in B} \|a - b\|, \max_{b \in B} \min_{a \in A} \|a - b\| \big)$$

where $\|\cdot\|$ denotes the underlying norm on $\mathbb{R}^d$, for example the Euclidean norm. So, the Hausdorff-distance considers the maximum distance of a point in one set to the other set and it gives reasonable results in many cases. Moreover, it can be computed efficiently in two dimensions if $A$ and $B$ consist of sets of line segments [ABB95] or, more generally, in arbitrary but fixed dimension $d$ where $A$ and $B$ consist of sets of $k$-dimensional simplices [ABG$^+$03].

However, if shapes are modeled by curves or surfaces, there are examples of objects having little resemblance but a small Hausdorff distance, see Figure 1.

[*]Freie Universität Berlin, Institut für Informatik, Takustr. 9, D-14195 Berlin, Germany. Email: {alt,mbuchin}@inf.fu-berlin.de. This research was partially supported by the Deutsche Forschungsgemeinschaft (DFG) within the framework of the graduate program (Graduiertenkolleg) "Combinatorics, Geometry, and Computation", No. GRK 588/3.

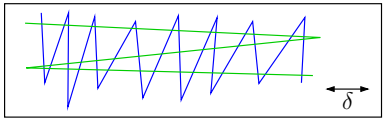

Figure 1: Nonsimilar polygonal chains with a small Hausdorff-distance $\delta$.

In these cases the *Fréchet distance* is more appropriate (see [AG95, AERW03]), which is a metric for parametrized geometric objects and was first introduced by Fréchet for curves [Fré06] and later for surfaces [Fré24]. The idea of the Fréchet distance is to take into account the "flow" of the curve or surface given by its parameterization. A popular illustration of the Fréchet distance of two curves is the following. Suppose a man is walking his dog on a leash. The man is walking on one curve and the dog on the other. Both may stop but not walk backwards. Then the Fréchet distance is the shortest length of leash allowing them to walk on the two curves from start to end.

Formally the Fréchet distance is defined as follows.

**Definition 1.** Let $f, g$ be parameterizations of curves or surfaces, i.e., continuous functions
$$f, g : [0,1]^k \to \mathbb{R}^d, \quad k \in \{1,2\}, \quad d \geq k.$$
Then their *Fréchet distance* is
$$\delta_F(f, g) = \inf_{\sigma:[0,1]^k \to [0,1]^k} \sup_{t \in [0,1]^k} \|f(t) - g(\sigma(t))\|.$$
where the *reparameterization* $\sigma$ ranges over all orientation preserving homeomorphisms.

In this paper, we assume that the norm $\|\cdot\|$ underlying the definition is the Euclidean norm but the results hold for the $L_1$-, and $L_\infty$-norm as well, and with some modifications can be generalized to other norms, such as any $L_p$-norm, $p \in \mathbb{N}$.

For dimension $k = 1$ of the parameter space, in particular for polygonal curves, the Fréchet distance is known to be computable in polynomial time [AG95]. For two-dimensional surfaces, however, the computation of the Fréchet distance surprisingly seems to be much harder. In fact, Godau showed in his Ph.D. thesis [God98] that computing the Fréchet distance between triangulated surfaces, even in two-dimensional space, is NP-hard. It remained open, however, how hard the problem really is, not even its computability could be shown.

For the special case that both surfaces are simple plane polygons (in a space of arbitrary dimension $d$) Buchin et al. [BBW06] showed that the Fréchet distance can be computed in polynomial time.

The first part of this paper contains a partial result concerning the computability of the Fréchet distance. More specifically, we will show that the Fréchet distance between triangulated surfaces is *upper semi-computable*, i.e., there is a non-halting Turing machine which produces a monotone decreasing sequence of rationals converging to the result. It follows that the decision problem whether the Fréchet distance of two given surfaces lies below some specified value is *recursively enumerable*.

The computationally hard part of computing the Fréchet distance for dimensions $k > 1$ seems to be, that according to the definition, the infimum over all homeomorphisms of the parameter space has to be taken. For dimension one the orientation-preserving homeomorphisms on $[0,1]$ are "only" the continuous, onto, monotone increasing functions on $[0,1]$. For higher dimensions the homeomorphisms can be much "wilder".

We tackle this problem in our algorithm by approximating the homeomorphisms by discrete maps which are easier to handle. We do this by first approximating arbitrary homeomorphisms by piecewise linear homeomorphisms which is a known result from topology. These piecewise linear homeomorphisms are then approximated by *mesh homeomorphisms*, i.e., ones that are compatible with certain subdivisions of the original triangulations of the parameter spaces. Finally, the distance obtained by considering only mesh homeomorphisms can be approximated for fine subdivisions by considering the distances at a finite number of points. It remains open, whether the Fréchet distance between triangulated surfaces is a computable function in the strong sense.

The second part of this paper considers a relaxed version of the Fréchet distance which we will call the *weak Fréchet distance* denoted by $\delta_{wF}$. It is defined by modifying Definition 1 so that the the reparameterization $\sigma$ of $g$ which is an orientation preserving homeomorphism is replaced by reparameterizations $\sigma, \tau$ of both surfaces which are *surjective* continuous maps. In the man-dog illustration for curves this means that both can choose arbitrary starting and ending points, can move forward and backward, and each one has to traverse his curve completely.

We will show that the weak Fréchet distance between two given triangulated surfaces can be computed in polynomial time. This is done by first considering the corresponding decision problem, i.e., the question, whether for given $f, g$, and $\varepsilon > 0$ the distance $\delta_{wF}(f, g) \leq \varepsilon$. The decision problem can be characterized by geometric properties of the so called *free space diagram*, a data structure originally introduced to solve the decision problem for polygonal curves [AG95]. This characterization leads to a polynomial time algorithm for the decision problem. Since it can be shown that the exact distance must be one of a finite set of critical values, we can use the decision routine in a search strategy to actually compute the weak Fréchet distance.

## 2 Semi-computability of the Fréchet distance

We assume that the input to our algorithm are two triangulated surfaces in space $\mathbb{R}^d, d \geq 2$, which are represented by *piecewise linear* parameterizations $f, g : [0,1]^2 \to \mathbb{R}^d$. For simplicity, we will denote the surfaces themselves by $f$ and $g$, as well.

Piecewise linear means that the parameter spaces of $f$ and $g$ are triangulated and on each triangle $\Delta = \langle u, v, w \rangle$, $f$ and $g$, respectively, are linear maps in the sense that $f(\lambda_1 u + \lambda_2 v + \lambda_3 w) = \lambda_1 f(u) + \lambda_2 f(v) + \lambda_3 f(w)$ for all $\lambda_1, \lambda_2, \lambda_3 \geq 0$ with $\lambda_1 + \lambda_2 + \lambda_3 = 1$ and $g$ has an analogous property.

We denote the triangulated parameter spaces of $f$ and $g$ by $K$ and $L$. The vertices of the individual triangles have rational coordinates, and the coefficients describing the linear maps are rational as well. Thus, a problem instance has a canonical finite representation which can be given as input to a Turing machine

(we will describe the algorithm in some high level language, however.)

We will show that the Fréchet distance between triangulated surfaces is computable in a weak sense according to the following definition which has been considered in the research community concerned with the computability and complexity of real functions (see, e.g., [WZ00]).

**Definition 2.** A function $\varphi : \mathbb{N} \to \mathbb{R}$ is called *upper (lower) semi-computable* iff there is a Turing machine which on input $x$ outputs an infinite, monotone decreasing (increasing) sequence of rational numbers converging to $\varphi(x)$.

Now we can formulate the first *main result* of this article:

**Theorem 1.** *The Fréchet distance between two triangulated surfaces in space $\mathbb{R}^d$, $d \geq 2$, is upper semi-computable.*

Theorem 1 immediately implies the following corollary, where $\langle f, g, a \rangle$ denotes some standard encoding of a triple consisting of two triangulated surfaces $f$ and $g$, and some rational $a > 0$.

**Corollary 1.** *The set $\{\langle f, g, a \rangle | \ \delta_F(f, g) < a\}$, i.e., the decision problem for the Fréchet distance is recursively enumerable.*

*Proof.* Consider the Turing machine producing a monotone decreasing sequence converging to $\delta_F(f, g)$ which exists by Theorem 1. Stop this Turing machine as soon as it produces a value less than $a$. Thus, the algorithm will eventually halt for all triples $\langle f, g, a \rangle$ in the language and it will run forever for the ones not in the language. □

The computability of $\delta_F$ in the strong sense of computability theory of real functions (see, e. g., [Wei00]) remains open, since the sequence produced by the algorithm proving Theorem 1 is not shown to *effectively* converge to $\delta_F(f, g)$. That is, we cannot compute an upper bound on the distance of the value produced by the algorithm after $k$ steps to the real value $\delta_F(f, g)$.

Also observe that Corollary 1 cannot be deduced from Theorem 1 any more, if we replace the <-sign in the definition of the set by a ≤-sign.

Our proof can be modified to show a weaker form of Theorem 1 for more *general surfaces*. More precisely, if we just assume that the parameterizations $f$ and $g$ are *computable real functions*, it is still correct that there is an algorithm producing on input $f, g$ (represented, say, by the Turing machines computing $f$ and $g$) an infinite sequence of rational numbers converging to $\delta_F(f, g)$. However, this sequence is not necessarily monotone decreasing, i.e., the corollary cannot be deduced any more.

## 2.1 Approximating the homeomorphisms

We will prove the semi-computability of the Fréchet distance by showing that homeomorphisms can be approximated arbitrarily closely by so-called mesh homeomorphisms, which are compatible with finite subdivisions of the parameter space.

First, let us recall some standard definitions and notations from topology. For a triangulation $T$ let $T^m$ denote its *m-th barycentric subdivision*, where in one subdivision step each triangle is subdivided into 6 triangles by the bisectors of its sides. By $|T|$ we denote the underlying space of $T$, i.e., the set of all points lying on triangles of $T$. In our case $|K| = |L| = [0,1]^2$. And by $mesh(T)$ we denote the maximal diameter of triangles in $T$ which gives a measure for the "fineness" of $T$.

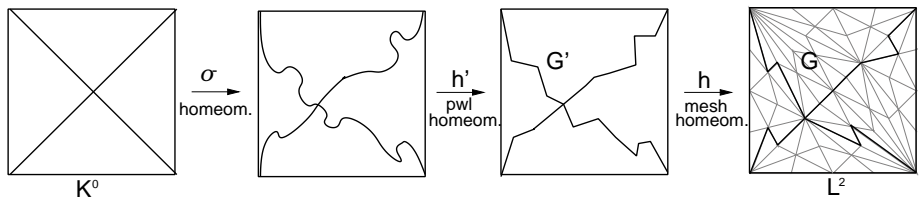

Figure 2: Approximating a homeomorphism by a mesh homeomorphism.

Now we define mesh homeomorphisms.

**Definition 3.** Given two triangulations $K$ and $L$, a piecewise linear homeomorphism $h : |K^m| \to |L^n|$ is called a *mesh homeomorphism* if it maps the edges of $K^m$ to *edge chains* of $L^n$, i.e., polygonal chains consisting of edges of $L^n$.

We will show that any homeomorphism can be approximated arbitrarily closely by a mesh homeomorphism. In fact, we need only a weak form of closeness which is defined as follows.

**Definition 4.** Let $K$ be a triangulation of the unit square and let $h, h' : [0,1]^2 \to [0,1]^2$ be homeomorphisms on the unit square. Then we define

$$d_K(h, h') = \max_{\Delta \in K} \delta_H(h(\Delta), h'(\Delta))$$

where $\Delta \in K$ ranges over all triangles in $K$ and $\delta_H$ denotes the Hausdorff distance.

**Lemma 1.** *Let $K$ and $L$ be triangulations, $\sigma : |K| \to |L|$ a homeomorphism, $m \in \mathbb{N}$, and $\varepsilon > 0$. Then there exist $n \in \mathbb{N}$ and a mesh homeomorphism $h : |K^m| \to |L^n|$ such that $d_{K^m}(\sigma, h) < \varepsilon$.*

*Proof.* By a theorem from topology (see e.g. chapter 6 in the book by Moise [Moi77]), $\sigma$ can be approximated arbitrarily closely by a piecewise linear homeomorphism, i.e., for all $\varepsilon_1 > 0$ there exists a piecewise linear homeomorphism $h' : |K| \to |L|$ with $d_{K^m}(\sigma, h') < \varepsilon_1$. We use this fact as a first step, because piecewise linear homeomorphisms are much easier to handle than arbitrary homeomorphisms.

We will show that any piecewise linear homeomorphism $h'$ can be approximated to any $\varepsilon_2 > 0$ (in the sense of Definition 4) by a mesh homeomorphism $h$, i.e., $d_{K^m}(h, h') < \varepsilon_2$. Choosing $\varepsilon_1$ and $\varepsilon_2$ so that $\varepsilon_1 + \varepsilon_2 < \varepsilon$ then proves the lemma, see Figure 2.

In order to approximate $h'$ we first show how to find edge chains in $L^n$, for some large enough $n \in \mathbb{N}$, that are close to the polygonal chains which are the images of edges of $K^m$ under $h'$. Then we explain how this can be extended to a piecewise linear homeomorphism on the whole parameter space $|K^m|$.

In fact, the piecewise linear images of the edges of $K^m$ under $h'$ form a graph $G'$ isomorphic to $K^m$ embedded in $|L|$ where the edges of $G'$ are polygonal chains

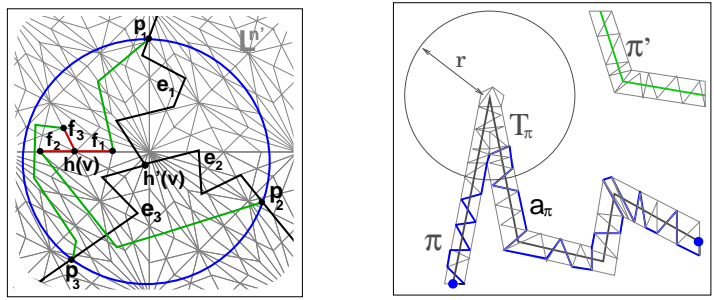

(a) Construction in the neighborhood of a vertex

(b) Edge chain $a_\pi$ approximating $\pi$

Figure 3: Approximating a piecewise linear homeomorphism by a mesh homeomorphism.

in $|L|$. We want to modify $G'$ to obtain an isomorphic graph $G$ embedded in $|L^n|$ with edge chains of $L^n$ as edges that have a distance smaller than $\varepsilon_2$ to the corresponding edges of $G'$.

We do this in two steps:

**Step 1.** We map the nodes of the graph $G'$ and short initial segments of their incident edges to nearby vertices of $L^{n'}$ and (some of) their incident edges, for some suitable $n' \in \mathbb{N}$.

More precisely, for mapping the nodes, we put a small circle of radius $r$ around each node $h'(v)$ where $r < \varepsilon_2/2$ and less than the smallest distance between two nodes $h'(v_1)$ and $h'(v_2)$. $r < \varepsilon_2/2$ yields disks of diameter less than $\varepsilon_2$ in which we may move nodes and edges freely. $r$ less than the smallest distance between two nodes ensures that disks around different nodes do not touch.

We choose $n' \in \mathbb{N}$ such that $mesh(L^{n'}) < r/2$ and for all nodes $h'(v)$ there is a vertex $w \in L^{n'}$ of distance less than $r/2$ whose degree (in $L^{n'}$) is greater than the degree of $h'(v)$ (in $G'$). This is possible because $mesh(L^{n'})$ tends to zero for large $n'$ and in the barycentric subdivision the degrees of vertices double in each subdivision step after their introduction. For each node $v \in K^m$ we choose $h(v) = w$ for such a $w$, i.e., satisfying $\|h'(v) - h(v)\| < r/2 < \varepsilon_2$ and $\deg(h(v)) \geq \deg(v)$. By this construction each $h(v)$ and all its incident edges of $L^{n'}$ lie in the disk of radius $r$ around $h'(v)$, see Figure 3 (a).

We start mapping the edges of $G'$ by first choosing edges incident to $h(v)$ in $L^{n'}$ as initial segments. We do so maintaining the order given by $G'$, i.e., if the edges $e_1, ..., e_k$ leave the vertex $h'(v)$ of $G'$ in clockwise order, we choose corresponding edges $f_1, ..., f_k$ of $L^{n'}$ leaving $h(v)$ in the same order (see Figure 3 a).

Next we cut the edges $e_1, ..., e_k$ at the points $p_1, ..., p_k$ where they first leave the disk. Within the disk we connect the free endpoint of each $f_i$ with the point $p_i$, $i = 1, ..., k$ by nonintersecting polygonal chains, which replace the original polygonal chains from $h'(v)$ to $p_i$. This is possible because we chose the edges $f_1, ..., f_k$ in the same order as the cutting points and within a disk we are allowed to move freely, without violating the distance bound $\varepsilon_2$ (see Figure 3 a). Thus, we replaced all vertices $h'(v)$ by closeby vertices $h(v)$ lying on the mesh $L^{n'}$.

**Step 2.** Next, we delete the nodes $h(v)$ together with the incident edges $f_1, ..., f_k$ from the scenery, leaving a finite set of pairwise disjoint polygonal chains which start and end in mesh points of $L^{n'}$. We show that they can be $\varepsilon_2$-approximated by edge chains of $L^n$, for suitable $n \geq n'$.

To achieve this, let $\eta$ be the minimum distance between a vertex of a curve and a nonincident edge of a curve (possibly the same). For edges sharing an incident vertex let $\theta$ be the minimum distance between the intersection points of the edges with a circle of radius $\varepsilon_2/2$ around their common vertex.

Then we subdivide $L^{n'}$ to $L^n$ for a suitable $n \geq n'$ such that $mesh(L^n) < \min(\eta/2, \theta/2)$. Thus, $L^n$ is so fine that each polygonal chain $\pi$ traverses a sequence of triangles $T_\pi$ so that $T_\pi$ and even its neighboring triangles are not intersected by other polygonal chains (see Figure 3 (b)).

Now, we can approximate each polygonal chain $\pi$ by any connected, not self-intersecting edge chain $a_\pi$ of $L^n$ that lies within $T_\pi$ and connects the endpoints of $\pi$ so that the Hausdorff distance between $\pi$ and $a_\pi$ is less than $\varepsilon_2$.

Thus, we showed the existence of a polygonal chain $h(e)$ in $L^n$ for each edge $e$ of $K^m$ which is arbitrarily close to $\sigma(e)$. The chains $h(e)$ and the vertices $h(v)$ form an embedded graph $G$ isomorphic to $G'$ and, therefore, to $K^m$. $h$ can be extended to the interior points of each edge $e$ to form a homeomorphism on $e$ in a straightforward manner. Furthermore, the faces of $G$ induce a partition of the set of triangles of $L^n$ which is isomorphic to the triangulation $K^m$. To extend $h$ to a piecewise linear homeomorphism on $|K^m|$, we subdivide each triangle $\Delta$ of $K^m$ according to the triangulation of the associated set of triangles in the partition of $L^n$ and extend $h$ to the interior of $\Delta$ correspondingly. $\square$

The fact, that homeomorphisms can be approximated arbitrarily closely by mesh homeomorphisms implies that the Fréchet distance can be approximated arbitrarily closely by considering only mesh homeomorphisms and distances between the vertices of the corresponding triangulations. More precisely, we obtain:

**Lemma 2.**

$$\delta_F(f,g) \;=\; \inf_{m,n \in \mathbb{N}} \; \inf_{h:|K^m| \to |L^n|} \; \max_{\Delta \in K_T^m} \; \max_{\substack{v \in V_\Delta \\ w \in M_{h(\Delta)}^n}} \; \|f(v) - g(w)\|$$

where $h$ ranges over all orientation preserving mesh homeomorphisms, $K_T^m$ is the set of triangles in $K^m$, $V_\Delta$ are the vertices of $\Delta$, and $M_{h(\Delta)}^n$ is the set of vertices of $L^n$ that lie in $h(\Delta)$.

*Proof.* In order to prove Lemma 2 we define the right hand side of the equation in the lemma as the *discrete Fréchet distance* $\delta_{dF}(f,g)$ of surfaces and we show that it is equal in value to the Fréchet distance.

We first show that the Fréchet distance is not larger than the discrete Fréchet distance.

**Claim 1.** $\delta_F \leq \delta_{dF}$.

Since any mesh homeomorphism is, in particular, a homeomorphism it suffices to show that for a mesh homeomorphism $h : |K^m| \to |L^n|$ we can bound the pointwise maximum by the maximum taken at vertices, i.e., show that

$$\max_{t \in [0,1]^2} \|f(t) - g(h(t))\| \leq \max_{\Delta \in K_T^m} \max_{\substack{v \in V_\Delta \\ w \in M_{h(\Delta)}^n}} \|f(v) - g(w)\|. \qquad (1)$$

To see this, let $t \in [0,1]^2$ be arbitrary. $t$ lies in a triangle $\Delta = \langle v_1, v_2, v_3 \rangle$ of $K^m$ and $h(t)$ lies in a triangle $\Delta' = \langle w_1, w_2, w_3 \rangle$ of $h(\Delta) \subset L^n$. Since $f$ and $g$ are piecewise linear and $K^m$ and $L^n$ are refinements of the underlying triangulations of the parameter spaces, $f(\Delta)$ and $g(\Delta')$ are triangles, as well, namely $\langle f(v_1), f(v_2), f(v_3) \rangle$ and $\langle g(w_1), g(w_2), g(w_3) \rangle$, respectively. Consequently, since the maximum distance between points of two triangles in 3-space is attained between two corners, we have that $\|f(t) - g(h(t))\| \leq \|f(v_i) - g(w_j)\|$ for some $i, j$ with $1 \leq i, j \leq 3$. Taking the maximum on both sides yields equation (1).

Now we show that also the discrete Fréchet distance is not larger than the Fréchet distance.

**Claim 2.** For all $\varepsilon > 0$, $\delta_{dF} \leq \delta_F + \varepsilon$.

The idea is that for any homeomorphism there is a mesh homeomorphism arbitrarily close and for the mesh homeomorphism the distance computation at vertices comes arbitrarily close to the distance computation on all parameter values by sufficient subdivision of the domain complex.

Let $\sigma$ be a homeomorphism close to realizing $\delta_F$, i.e., $\max_t \|f(t) - g(\sigma(t))\| \leq \delta_F + \varepsilon_1$ for some small $\varepsilon_1 > 0$. By Lemma 1, for any $\varepsilon_2 > 0$ and any $m \in \mathbb{N}$ there is a mesh homeomorphism $h : |K^m| \to |L^n|$ such that $d_{K^m}(\sigma, h) \leq \varepsilon_2$.

Let $\Delta$ be some triangle in $|K^m|$ and $v$ one of its vertices. Since $d_{K^m}(\sigma, h) \leq \varepsilon_2$, for any $w \in h(\Delta) \subset L^n$ there is an $x \in \sigma(\Delta)$ with $\|w - x\| < \varepsilon_2$. Using $t = \sigma^{-1}(x)$ and the Lipschitz-continuity of $g$ we get $\|g(w) - g(\sigma(t))\| < c_g \cdot \varepsilon_2$ for some $t \in \Delta$ where $c_g$ denotes the Lipschitz constant of $g$.

$t$ and $v$ lie in the same triangle $\Delta \in K^m$, so $\|v - t\| \leq \mathrm{mesh}(K^m)$ and, thus, $\|f(v) - f(t)\| \leq c_f \cdot \mathrm{mesh}(K^m)$ where $c_f$ is the Lipschitz constant for $f$.

Putting everything together and applying the triangle inequality repeatedly we obtain

$$\begin{aligned}
\delta_{dF} &\leq \max_{\Delta \in K_T^m} \max_{\substack{v \in V_\Delta \\ w \in M_{h(\Delta)}^n}} \|f(v) - g(w)\| \\
&\leq \max_{\Delta \in K_T^m} \max_{\substack{v \in V_\Delta \\ x \in \sigma(\Delta)}} \|f(v) - g(x)\| + c_g \cdot \varepsilon_2 \\
&\leq \max_{\Delta \in K_T^m} \max_{t \in \Delta} \|f(t) - g(\sigma(t))\| + c_g \cdot \varepsilon_2 + c_f \cdot \mathrm{mesh}(K^m) \\
&\leq \delta_F + \varepsilon_1 + c_g \cdot \varepsilon_2 + c_f \cdot \mathrm{mesh}(K^m).
\end{aligned}$$

Since $\varepsilon_1, \varepsilon_2$, and, by choosing $m$ large enough, $\mathrm{mesh}(K^m)$ can be made arbitrarily small, Claim 2 is shown.

Claim 1 and 2 yield $\delta_F = \delta_{dF}$ which proves Lemma 2. $\square$

## 2.2 Semi-computing the Fréchet distance

Using Lemma 2 we can now give an algorithm showing the upper semi-computability of the Fréchet distance between surfaces as claimed in Theorem 1. This algorithm will, on input $f, g$ run forever and produce a monotone decreasing sequence of rational numbers converging to $\delta_F(f, g)$.

---
**Algorithm 1**: SemiComputeFrechet($f, g$)

---
**Input**: Triangulated surfaces $f, g$, including triangulations $K, L$ of the parameter spaces, in a finite description
**Output**: A monotone decreasing sequence of rational numbers converging to $\delta_F(f, g)$

1 set $D = \infty$
2 **forall** $(n, m) \in \mathbb{N} \times \mathbb{N}$ **do**
3     generate the barycentric subdivisions $K^m$ of $K$ and $L^n$ of $L$
4     let $E = \{e_1, ..., e_k\}$ be the set of edges in $K^m$
5     **forall** $k$-tuples $(\pi_1, ..., \pi_k)$ of simple polygonal chains in $L^n$ **do**
6        assign to the edge $e_i$ the polygonal chain $\pi_i$ for $i = 1, ..., k$
7        **if** *this assignment results in an orientation preserving homeomorphic image of $K^m$* **then**
8            set $M = 0$
9            **forall** *triangles $\Delta$ of $K^m$* **do**
10               let $H_\Delta \subset |L^n|$ be the region in $L^n$ assigned to $\Delta$
11               **forall** *vertices $v$ of $\Delta$ and vertices $w$ of $H_\Delta$* **do**
12                  set $M = \max(M, \|f(v) - g(w)\|)$
13               **end**
14            **end**
15            set $D = \min(D, M)$
16            output $D$
17        **end**
18     **end**
19 **end**

---

Line 2 can be realized by some standard enumeration method for pairs of integers. The number of $k$-tuples of polygonal chains of $L^n$ checked in line 5 is finite. In fact, it is bounded by $(l!)^k$ where $l$ is the number of edges in $L^n$, which itself is exponential in $n$, whereas $k$ is exponential in $m$. But efficiency is not the issue here.

In line 12 we assume that the norm $\|\cdot\|$ underlying the Fréchet distance can be evaluated by rational operations. This is correct for, e.g., the $L_1$- or $L_\infty$-metric but not directly for the Euclidean metric $L_2$. In that case, one should rather operate with the square of the distance in line 12 and output some suitable rational approximation of $\sqrt{D}$ (which is possible) in line 16.

In line 7 we check whether an assignment of edges in $K^m$ to polygonal chains in $L^n$ results in an orientation-preserving homeomorphic image of $K^m$ by checking the following three conditions

- the edges on the boundary of $|K^m|$ are mapped onto the boundary of $|L^n|$ preserving the orientation

- if a set of edges in $K^m$ share an endpoint, the corresponding chains in $L^n$

do, as well,

- other than that, there are no intersection points between two chains.

Note that checking that the boundary of $|K^m|$ is mapped orientation preservingly onto the boundary of $|L^n|$ entails that the mesh homeomorphism is orientation preserving also on the interior.

For each pair $(m,n) \in \mathbb{N} \times \mathbb{N}$ all mesh homeomorphisms $h: K^m \to L^n$ are evaluated by Algorithm 1, i.e.,

$$\delta_{h,m,n} = \max_{\Delta \in K_T^m} \max_{\substack{v \in V_\Delta \\ w \in M_{h(\Delta)}^n}} \|f(v) - g(w)\|$$

(see Lemma 2) is computed[1].

To see that Algorithm 1 produces values arbitrarily close to $\delta_F(f,g)$, observe that any neighborhood of $\delta_F(f,g)$ must, by Lemma 2, contain some value of the form $\delta_{h,m,n}$. The algorithm will eventually encounter that pair $(m,n)$ and the subdivision corresponding to $h$ and output $\delta_{h,m,n}$. By line 15 the output sequence is monotone decreasing. Since for all triples $(h,m,n)$, by Lemma 2, $\delta_{h,m,n} \geq \delta_F(f,g)$, line 15 is justified. Therefore, Algorithm 1 arbitrarily closely approximates $\delta_F(f,g)$ which proves Theorem 1.

## 3   Computability of the Weak Fréchet Distance

In this section we give a polynomial time algorithm for computing the weak Fréchet distance between triangulated surfaces. The weak Fréchet distance is a relaxed version of the Fréchet distance which does not require the reparameterizations of the curves or surfaces to be injective. Instead it uses surjective continuous maps as reparameterizations.

**Definition 5.** The weak Fréchet distance between curves or surfaces given by continuous parameterizations $f, g: [0,1]^k \to \mathbb{R}^d$ with $k \in \{1,2\}, k \leq d$ is

$$\delta_{wF}(f,g) = \inf_{\alpha, \beta: [0,1]^k \to [0,1]^k} \max_{\mathbf{x} \in [0,1]^k} \|f(\alpha(\mathbf{x})) - g(\beta(\mathbf{x}))\| ,$$

where $\alpha$ and $\beta$ range over all surjective continuous maps.

As was mentioned before in the man-dog illustration for curves this means that both are allowed to choose their start and endpoints, can move forward and backward, and eventually must have traversed both curves completely. The weak Fréchet distance is always less than or equal to the Fréchet distance and greater or equal the Hausdorff distance:

$$\delta_H(f,g) \leq \delta_{wF}(f,g) \leq \delta_F(f,g).$$

Figure 4, e.g., shows two curves with a small weak Fréchet distance and a large Fréchet distance. For curves, this relation between the distance measures

---

[1] A more detailed analysis shows that, in fact, it suffices to consider only the pairs $(m, 2m), m \in \mathbb{N}$

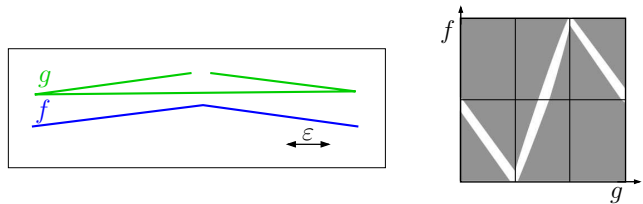

Figure 4: Curves with small small weak Fréchet distance and their free space diagram for parameter $\varepsilon$

is reflected in the *free space diagram* which is a geometric data structure for computing the Fréchet distance of polygonal curves [AG95].

For the sake of illustration we will shortly repeat these ideas and concepts for curves before we extend them to surfaces. For two parameterized curves $f, g$ and parameter $\varepsilon$ the *free space* is defined as $F_\varepsilon(f,g) = \{(x,y) \mid \|f(x) - g(y)\| \leq \varepsilon\}$. The free space diagram illustrates the free space, see Figure 4, where the direction from left to right corresponds to the parameterization of $f$ and the one from bottom to top to the parameterization of $g$. The white area inside the diagram corresponds to the free space for the indicated value of $\varepsilon$. For polygonal curves $f, g$ of lengths $n$ and $m$, respectively, the free space diagram consists of $nm$ cells, each one corresponding to the free space between two segments of the curves.

It can easily be seen, that $\delta_F(f,g) \leq \varepsilon$ if and only if there is a monotone path in the free space from the lower left corner $(0,0)$ to the upper right corner $(1,1)$ (corresponding to the reparameterization $\sigma$ in Definition 1). This fact is used in [AG95] to compute the Fréchet distance.

But also the fact that the Hausdorff distance or the weak Fréchet distance are less than or equal $\varepsilon$ can be characterized by the free space diagram. Let us call the free space or any subset of it *extensive* if its projections onto both parameter spaces are surjective. Then:

Hausdorff distance less than or equal $\varepsilon$ is equivalent to the free space being extensive.

Weak Fréchet distance less than or equal $\varepsilon$ means that there is a curve inside the free space which is extensive. This fact is used in [AG95] to compute a restricted version of the weak Fréchet distance called the *non-monotone Fréchet distance* of polygonal curves. An extensive curve exists exactly if there is an extensive connected component of the free space.

Essentially, this concept and this characterization of the weak Fréchet distance can be extended to surfaces (yielding a four-dimensional free space diagram). We will use this idea to find an efficient algorithm for computing the weak Fréchet distance of triangulated surfaces.

In fact, the second *main result* of this article is the following theorem.

**Theorem 2.** *The weak Fréchet distance between triangulated surfaces can be computed in $O((m^2n + mn^2)\log^2(mn))$ time.*

The algorithms we give will need to compute the intersection of ellipses, circles and line segments and compare such intersection points. These intersection points can all be described as roots of polynomials of degree up to four, which can be compared and computed exactly using constantly many arithmetic operations on rationals, see [ET04, MPS$^+$06].

## 3.1 Free Space Diagram of Triangulated Surfaces

As in Section 2, we assume we are given piecewise linear parameterizations $f, g : [0,1]^2 \to \mathbb{R}^d$, $d \geq 2$, of triangulated surfaces with underlying triangulations $K, L$, respectively. All coefficients of the maps and coordinates of the triangulations are assumed to be rational. In the following, we will use $\mathbf{x}, \mathbf{y}$ to denote points in parameter space, i.e., $\mathbf{x}, \mathbf{y} \in [0,1]^2$. Let $F_\varepsilon(f, g) = \{(\mathbf{x}, \mathbf{y}) \mid \|f(\mathbf{x}) - g(\mathbf{y})\| \leq \varepsilon\}$ be the free space diagram of the surfaces. The free space diagram lies in the product of the parameter spaces of the surfaces, i.e., in the four-dimensional cube. As in the case of curves, we can partition the free space into cells.

**Definition 6.** Let $f, g : [0,1]^2 \to \mathbb{R}^d$, $d \geq 2$, be piecewise linear surface parameterizations with underlying triangulations $K$ and $L$, respectively. A *cell* of the free space is $C_\varepsilon(\Delta_K, \Delta_L) := F_\varepsilon(f, g) \cap (\Delta_K \times \Delta_L)$ for triangles $\Delta_K$ of $K$ and $\Delta_L$ of $L$. A *boundary cell* is the cell of a triangle and an edge, defined analogously. Two cells are called *neighboring* if they share a nonempty boundary cell.

We will consider the projection of the free space onto the two parameter spaces, i.e., the image of $F_\varepsilon(f, g)$ under $\mathrm{Proj}_K \colon [0,1]^4 \to [0,1]^2$, $(\mathbf{x}, \mathbf{y}) \mapsto \mathbf{x}$, and $\mathrm{Proj}_L \colon [0,1]^4 \to [0,1]^2$, $(\mathbf{x}, \mathbf{y}) \mapsto \mathbf{y}$. We will compute the projections of the free space by computing it for its cells, i.e., we will compute

$$\mathrm{Proj}_K\bigl(C_\varepsilon(\Delta_K, \Delta_L)\bigr) \quad = \quad \bigl\{ \mathbf{x} \in \Delta_K \mid \exists \mathbf{y} \in \Delta_L : \|f(\mathbf{x}) - g(\mathbf{y})\| \leq \varepsilon \bigr\},$$

and analogously for the projection onto the parameter space $L$. Although the free space lies in the product of the parameter spaces, it is defined based on the distances in image space.

The combinatorial structure of the free space is captured by the graph whose vertices are the cells of the free space and edges exist between neighboring cells which share a non-empty boundary cell. For a free space diagram $F_\varepsilon(f, g)$ we denote the corresponding graph by $G_\varepsilon$. The graph $G_\varepsilon$ has $mn$ vertices and $O(mn)$ edges. For $\varepsilon_1 \leq \varepsilon_2$ the graph $G_{\varepsilon_1}$ is a subgraph of $G_{\varepsilon_2}$. For varying $\varepsilon$, $G_\varepsilon$ changes only at finitely many values, namely when a boundary cell becomes non-empty. This happens when $\varepsilon$ equals the minimum distance between points on some edge $e_K$ of $K$ and on some triangle $\Delta_L$ of $L$ or vice versa:

$$C_\varepsilon(e_K, \Delta_L) \neq \emptyset \quad \Leftrightarrow \quad min\bigl\{\|x - y\| \mid x \in f(e_K), y \in g(\Delta_L)\bigr\} \leq \varepsilon. \quad (2)$$

The main idea of our algorithm is contained in the following lemma.

**Lemma 3.** *The weak Fréchet distance between two triangulated surfaces is less than or equal $\varepsilon$ if and only if there is an extensive connected component $A$ in the free space for the parameter $\varepsilon$, i.e., $Proj_K(A) = Proj_L(A) = [0,1]^2$.*

*Proof.* Let $\alpha, \beta$ be two reparameterizations of $f$ and $g$, respectively. We define a set $M_{\alpha, \beta} \subset [0,1]^4$ by

$$M_{\alpha, \beta} = \bigl\{ \bigl(\alpha(\mathbf{x}), \beta(\mathbf{x})\bigr) \mid \mathbf{x} \in [0,1]^2 \bigr\}.$$

The set $M_{\alpha,\beta}$ is connected because $\alpha$ and $\beta$ are continuous. $M_{\alpha,\beta}$ is extensive because $\alpha$ and $\beta$ are surjective. By definition of $M_{\alpha,\beta}$ the following equivalence holds:
$$M_{\alpha,\beta} \subseteq F_\varepsilon(f,g) \Leftrightarrow \max_{\mathbf{x}\in[0,1]^2} \|f(\alpha(\mathbf{x})) - g(\beta(\mathbf{x}))\| \leq \varepsilon. \qquad (3)$$

Now we show the two directions of the lemma.

$\Rightarrow$: If $\alpha, \beta$ are two reparameterizations for $f$ and $g$, respectively, realizing weak Fréchet distance $\leq \varepsilon$, i.e., $\max_{\mathbf{x}\in[0,1]^2}\|f(\alpha(\mathbf{x})) - g(\beta(\mathbf{x}))\| \leq \varepsilon$ then $M_{\alpha,\beta}$ is an extensive, connected subset of $F_\varepsilon(f,g)$, so the connected component containing $M_{\alpha,\beta}$ is an extensive connected component in $F_\varepsilon(f,g)$.

A weak Fréchet distance of at most $\varepsilon$, however, does not imply that two realizing reparameterizations $\alpha, \beta$ exist, as it is defined as $\inf_{\alpha,\beta} \max_{(\mathbf{x})} \|f(\alpha(\mathbf{x})) - g(\beta(\mathbf{x}))\| \leq \varepsilon$. This can be reformulated as
$$\forall \vartheta > \varepsilon \;\exists \alpha, \beta: \;\max_{\mathbf{x}\in[0,1]^2}\|f(\alpha(\mathbf{x})) - g(\beta(\mathbf{x}))\| \leq \vartheta.$$

By equation 3 this implies that $F_\vartheta(f,g)$ contains for all $\vartheta > \varepsilon$ an extensive connected component. We will show that this property holds also for $F_\varepsilon(f,g)$.

As discussed in Section 3.1 the combinatorial structure of the connected components of the free space changes only at finitely many values for $\varepsilon$, i.e., when a boundary cell becomes non-empty. Let

$\eta = \arg\min\{\vartheta \mid \vartheta > \varepsilon$ and a boundary cell becomes non-empty in $F_\vartheta(f,g)\}$.

Then the combinatorial structure of $F_\vartheta$ is the same for all $\vartheta \in [\varepsilon, \eta)$.

Let $(\varepsilon_n)_{n\in\mathbb{N}}$ be a sequence in $[\varepsilon, \eta)$ converging to $\varepsilon$. $F_{\varepsilon_n}$ contains for all $n$ an extensive connected component. A free space diagram has only finitely many connected components, and therefore there must be at least one extensive connected component in infinitely many of the $F_{\varepsilon_n}$. We claim that this connected component is also extensive in $F_\varepsilon$. Assume this is not the case, i.e., there is a point in parameter space that does not lie in the projection of the connected component in $F_\varepsilon$. Then because the projection of the free space is a closed set this must have already been the case for a small neighborhood $[\varepsilon, \eta') \subset [\varepsilon, \eta)$, i.e., for all but finitely many of the $\varepsilon_n, n \in \mathbb{N}$. This is a contradiction to the assumption.

$\Leftarrow$: Let $A \subset F_\varepsilon(f,g)$ be an extensive connected component of the free space. We will construct reparameterizations $\alpha$ and $\beta$ of $f$ and $g$, respectively, such that $M_{\alpha,\beta} \subset A$ and $\text{Proj}_K(M_{\alpha,\beta}) = \text{Proj}_K(A) = [0,1]^2$ and $\text{Proj}_L(M_{\alpha,\beta}) = \text{Proj}_L(A) = [0,1]^2$. This will imply that $\alpha$ and $\beta$ realize a weak Fréchet distance less than $\varepsilon$.

We do this in two steps: first we define for each cell $C_\varepsilon(\Delta_K, \Delta_L)$ contained in the connected component $A$ two parameterized surface patches $M_K(\Delta_K, \Delta_L)$ and $M_L(\Delta_K, \Delta_L)$, which fulfill $\text{Proj}_K(C_\varepsilon(\Delta_K, \Delta_L)) = \text{Proj}_K(M_K(\Delta_K, \Delta_L))$ and $\text{Proj}_L(C_\varepsilon(\Delta_K, \Delta_L)) = \text{Proj}_L(M_L(\Delta_K, \Delta_L))$. By a *surface patch* we mean a two-dimensional manifold with boundary in $\mathbb{R}^4$. Then we show how to "glue" two such parameterized surfaces patches together. Glueing together the surface patches of all cells contained in the connected component will give the desired reparameterizations.

**Defining Parameterized Surface Patches in each Cell**  For each cell $C_\varepsilon(\Delta_K, \Delta_L)$ of the free space we define two surface patches:

$$M_K(\Delta_K, \Delta_L) = \{(\mathbf{x}, g^{-1}(n(f(\mathbf{x}), g(\Delta_L)))) \mid \mathbf{x} \in \text{Proj}_K(C_\varepsilon(\Delta_K, \Delta_L))\},$$
$$M_L(\Delta_K, \Delta_L) = \{(f^{-1}(n(g(\mathbf{y}), f(\Delta_K)), \mathbf{y})) \mid \mathbf{y} \in \text{Proj}_L(C_\varepsilon(\Delta_K, \Delta_L))\},$$

where

$$n(x, A) = \arg\min_{y \in A} \|x - y\|$$

is the nearest neighbor of a point $x$ in the set $A$.

These surface patches have the following properties:

1. $M_K(\Delta_K, \Delta_L) \cup M_L(\Delta_K, \Delta_L) \subset C_\varepsilon(\Delta_K, \Delta_L)$.

2. $\text{Proj}_K(C_\varepsilon(\Delta_K, \Delta_L)) = \text{Proj}_K(M_K(\Delta_K, \Delta_L))$ and
   $\text{Proj}_L(C_\varepsilon(\Delta_K, \Delta_L)) = \text{Proj}_L(M_L(\Delta_K, \Delta_L))$.

3. $M_K(\Delta_K, \Delta_L) \cap M_L(\Delta_K, \Delta_L) \neq \emptyset$.

4. If the shared boundary cell of the cells $C_\varepsilon(\Delta_K, \Delta_L)$ and $C_\varepsilon(\Delta_K, \Delta'_L)$ is non-empty, then $M_L(\Delta_K, \Delta_L) \cap M_L(\Delta_K, \Delta'_L) \neq \emptyset$.
   If the shared boundary cell of the cells $C_\varepsilon(\Delta_K, \Delta_L)$ and $C_\varepsilon(\Delta'_K, \Delta_L)$ is non-empty, then $M_K(\Delta_K, \Delta_L) \cap M_K(\Delta'_K, \Delta_L) \neq \emptyset$.

5. Parameterizations $\alpha_K, \beta_K$ and $\alpha_L, \beta_L$ exist which fulfill $M_K(\Delta_K, \Delta_L) = M_{\alpha_K, \beta_K}$ and $M_L(\Delta_K, \Delta_L) = M_{\alpha_L, \beta_L}$.

For the first property consider a tuple $(\mathbf{x}, \mathbf{y}) \in M_K(\Delta_K, \Delta_L)$. By definition the point $\mathbf{x}$ is in $\text{Proj}_K(C_\varepsilon(\Delta_K, \Delta_L))$, i.e., there is a point $\mathbf{y}'$ s.t. $(\mathbf{x}, \mathbf{y}') \in C_\varepsilon(\Delta_K, \Delta_L)$. But then also $(\mathbf{x}, \mathbf{y}) \in C_\varepsilon(\Delta_K, \Delta_L)$ because $\mathbf{y}$ was chosen based on the nearest neighbor in image space.

The second property holds by definition, that is, $M_K(\Delta_K, \Delta_L)$ is defined by choosing a tuple for each point $\mathbf{x} \in \text{Proj}_K(C_\varepsilon(\Delta_K, \Delta_L))$.

The third property is equivalent to the existence of two points in $\Delta_K$ and $\Delta_L$ whose images under $f$ and $g$, respectively, are each others nearest neighbor, i.e., a tuple $(\mathbf{x}, \mathbf{y}) \in \Delta_K \times \Delta_L$ such that $g(\mathbf{y}) = n(f(\mathbf{x}), g(\Delta_L))$ and $f(\mathbf{x}) = n(g(\mathbf{y}), f(\Delta_K))$. Points fulfilling this condition are points whose images have minimal distance in the free space cell, i.e., a tuple $(\mathbf{x}, \mathbf{y})$ such that $\|f(\mathbf{x}) - g(\mathbf{y})\| \leq \|f(\mathbf{x}') - g(\mathbf{y}')\|$ for all tuples $(\mathbf{x}', \mathbf{y}') \in \Delta_K \times \Delta_L$. Such a tuple exists because $\Delta_K \times \Delta_L$ is compact.

The fourth property holds because $M_L(\Delta_K, \Delta_L)$ and $M_L(\Delta_K, \Delta'_L)$ coincide on the intersection of $C_\varepsilon(\Delta_K, \Delta_L)$ and $C_\varepsilon(\Delta_K, \Delta'_L)$, and analogously for $M_K(\Delta_K, \Delta_L)$ and $M_K(\Delta'_K, \Delta_L)$.

The fifth property is fulfilled by the following natural parameterizations. Let $\sigma_K$ and $\sigma_L$ be homeomorphisms from $[0, 1]^2$ to $\Delta_K$ and $\Delta_L$, respectively. Then $\alpha_K = \sigma_K$ and $\beta_K = g^{-1} \circ n(\cdot, g(\Delta_L)) \circ f \circ \sigma_K$ and $\alpha_L = f^{-1} \circ n(\cdot, f(\Delta_K)) \circ g \circ \sigma_L$ and $\beta_L = \sigma_L$ are parameterizations of $M_K(\Delta_K, \Delta_L)$ and $M_L(\Delta_K, \Delta_L)$, respectively. These reparameterizations are continuous because the identity map, $f, g$ and the nearest neighbor function on convex sets in $\mathbb{R}^d$ are continuous.

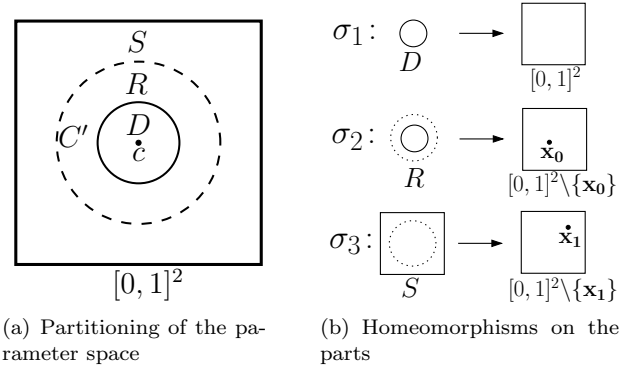

(a) Partitioning of the parameter space

(b) Homeomorphisms on the parts

Figure 5: Glueing together two parameterizations

Now we define $M_A$ to be the union of all surface patches of cells in the connected component $A$, i.e.,

$$M_A = \bigcup_{C_\varepsilon(\Delta_K, \Delta_L) \subset A} \bigl(M_K(\Delta_K, \Delta_L) \cup M_L(\Delta_K, \Delta_L)\bigr).$$

The first two properties of the surface patches $M_K(\Delta_K, \Delta_L)$ and $M_L(\Delta_K, \Delta_L)$ imply the following properties of $M_A$:

1. $M_A \subset A$

2. $\text{Proj}_K(A) = \text{Proj}_K(M_A)$ and $\text{Proj}_L(A) = \text{Proj}_L(M_A)$.

**Glueing together the Parameterizations of the Surface Patches** We first show that two reparameterizations can be joined to one and then obtain parameterizations $\alpha_A, \beta_A$ for $M_A$ by successively glueing together parameterizations of the surface patches in $M_A$.

**Claim 3.** Given parameterizations $\alpha, \beta, \alpha', \beta'$ with $M_{\alpha,\beta} \cap M_{\alpha',\beta'} \neq \emptyset$. Then there exist parameterizations $\alpha'', \beta''$ such that $M_{\alpha,\beta} \cup M_{\alpha',\beta'} = M_{\alpha'',\beta''}$.

*Proof of Claim 3:* Let $\mathbf{p} = (\mathbf{x_p}, \mathbf{y_p})$ be an intersection point of $M_{\alpha,\beta}$ and $M_{\alpha',\beta'}$. Let $\mathbf{x_0}, \mathbf{x_1}$ be such that $\mathbf{p} = \bigl(\alpha(\mathbf{x_0}), \beta(\mathbf{x_0})\bigr)$ and $\mathbf{p} = \bigl(\alpha'(\mathbf{x_1}), \beta'(\mathbf{x_1})\bigr)$.

For glueing together the parameterizations $\alpha, \beta, \alpha', \beta'$, we partition the parameter space as illustrated in Figure 5 (a). Choose a point $\mathbf{c}$ in parameter space and radii $r, r'$ such that $r < r'$ and the disk of radius $r'$ around the point $\mathbf{c}$ is contained in $[0,1]^2$. Let $C, C'$ be the circles of radius $r, r'$ around $\mathbf{c}$ and $D, D'$ the (closed) disks of radius $r, r'$ around $\mathbf{c}$. Let $R$ be the half-open ring with radii $r, r'$ around $\mathbf{c}$ and let $S = [0,1]^2 \setminus D'$.

We will glue the parameterizations $\alpha, \beta$ and $\alpha', \beta'$ along the circle $C'$. On the circle $\alpha'', \beta''$ will be constant, outside they will be defined based on $\alpha', \beta'$ and inside based on $\alpha, \beta$.

We define homeomorphisms on the parts of the partition as illustrated in Figure 5 (b). Let $\sigma_1$ be a homeomorphism from $D$ to the unit cube. Let $\sigma_2$ be a homeomorphism from $R$ to the unit cube without the point $\mathbf{x_0}$. Choose $\sigma_1$ and $\sigma_2$ s.t. they coincide on the circle $C$. Let $\sigma_3$ be a homeomorphism from $S$ to $[0,1]^2 \setminus \{\mathbf{x_1}\}$. Now we define $\alpha'', \beta'' : [0,1]^2 \to [0,1]^2$ as follows:

$$\alpha''(\mathbf{x}) = \begin{cases} \alpha(\sigma_1(\mathbf{x})), & \mathbf{x} \in D \\ \alpha(\sigma_2(\mathbf{x})), & \mathbf{x} \in R \\ \mathbf{x_p}, & \mathbf{x} \in C' \\ \alpha'(\sigma_3(\mathbf{x})), & \mathbf{x} \in S \end{cases}, \qquad \beta''(\mathbf{x}) = \begin{cases} \beta(\sigma_1(\mathbf{x})), & \mathbf{x} \in D \\ \beta(\sigma_2(\mathbf{x})), & \mathbf{x} \in R \\ \mathbf{y_p}, & \mathbf{x} \in C' \\ \beta'(\sigma_3(\mathbf{x})), & \mathbf{x} \in S \end{cases}.$$

$\alpha'', \beta''$ are continuous and fulfill $M_{\alpha,\beta} \cup M_{\alpha',\beta'} = M_{\alpha'',\beta''}$ by construction. They are continuous inside $D, R$, and $D$ as concatenations of continuous functions. On the circle $C$ they are continuous because $\sigma_1$ and $\sigma_2$ coincide on $C$. They are continuous on the circle $C'$ because $\sigma_2(\mathbf{x})$ tends to $\mathbf{x_0}$ for $\mathbf{x}$ tending to $\mathbf{x}' \in C'$ and therefore $\alpha(\sigma_2(\mathbf{x}))$ and $\beta(\sigma_2(\mathbf{x}))$ tend to $\mathbf{x_p}$ and $\mathbf{y_p}$, respectively, for $\mathbf{x}$ tending to $\mathbf{x}' \in C'$. Similarly, $\sigma_3(\mathbf{x})$ tends to $\mathbf{x_1}$ for $\mathbf{x}$ tending to $\mathbf{x}' \in C'$ and therefore $\alpha'(\sigma_3(\mathbf{x}))$ and $\beta'(\sigma_3(\mathbf{x}))$ tend to $\mathbf{x_p}$ and $\mathbf{y_p}$, respectively, for $\mathbf{x}$ tending to $\mathbf{x}' \in C'$. Thus, Claim 3 is proven.

We can glue together all surface patches in $M_A$ because they are connected by intersection – properties 3 and 4 of the surface patches. By glueing together all surface patches in $M_A$ we get reparameterizations $\alpha_A, \beta_A$ for $f, g$ which realize a weak Fréchet distance between $f$ and $g$ which is less than $\varepsilon$: $\alpha_A$ and $\beta_A$ are continuous because the parameterizations of single surface patches are continuous and the glueing of parameterizations preserves continuity. $\alpha_A$ and $\beta_A$ are surjective because the projection of $M_{\alpha_A, \beta_A} = M_A$ equals the projection of $A$ – property 2 of $M_A$ – and this was assumed to be $[0,1]^2$. Finally, $\|f(\alpha_A(\mathbf{x})) - g(\beta_A(\mathbf{x}))\| \leq \varepsilon$ holds for all $\mathbf{x} \in [0,1]^2$ because $M_{\alpha_A, \beta_A}$ lies completely in $F_\varepsilon(f,g)$. $\square$

## 3.2 Decision Problem for the Weak Fréchet distance

By Lemma 3, deciding if the weak Fréchet distance is less than some given value $\varepsilon$ is equivalent to determining if there is an extensive connected component in the free space for the parameter $\varepsilon$. We will do this algorithmically in two steps: First we compute all connected components of the free space, secondly we test whether there is an extensive one.

For the first step we compute the combinatorial graph of the free space as described in the previous section in $O(mn)$ time.

In the second step we need to determine if the projection of a connected component $A$ completely covers both parameter spaces. This property for, e.g., the parameter space $K$ is is equivalent to each triangle $\Delta_K$ in $K$ being covered by the projection of $A$. Equivalently, $f(\Delta_K)$ in image space must be completely contained in the union of the $\varepsilon$–neighborhoods $g(\Delta_L) \oplus B_\varepsilon$ of all triangles $\Delta_L$ such that the cell $C_\varepsilon(\Delta_K, \Delta_L)$ lies in $A$.

We can reduce this to a two-dimensional problem by intersecting the $\varepsilon$–neighborhoods $g(\Delta_L) \oplus B_\varepsilon$ with the plane $E_{\Delta_K}$ which contains the triangle $f(\Delta_K)$. I.e., we decide for each triangle $\Delta_K$ in $K$ whether

$$f(\Delta_K) \subset \bigcup_{\substack{\Delta_L \in L \\ C_\varepsilon(\Delta_K, \Delta_L) \in A}} \Big( (g(\Delta_L) \oplus B_\varepsilon) \cap E_{\Delta_K} \Big). \tag{4}$$

This property can be decided by computing for each triangle $\Delta_K$ the arrangement of the triangle $f(\Delta_K)$ and the set of boundary curves $\partial\big((g(\Delta_L) \oplus B_\varepsilon)\big) \cap E_{\Delta_K})$ for all triangles $\Delta_L \in L$ with $C_\varepsilon(\Delta_K, \Delta_L) \in A$ where $\partial$ denotes the boundary operator. Then we apply a line sweep algorithm to the part of the arrangement contained in the triangle $f(\Delta_K)$. Property (4) holds if and only if there is no empty cell in this part of the arrangement.

For the metrics $L_1$ and $L_\infty$, the $\varepsilon$-neighborhoods can be described by linear equations. For the Euclidean metric the $\varepsilon$-neighborhoods are described by quadratic equations: the $\varepsilon$-neighborhood of a triangle in 3-space is the union of $\varepsilon$-balls around the vertices, cylinders of radius $\varepsilon$ around the edges and a triangular prism of height $\varepsilon$. The boundaries of the intersections of these $\varepsilon$-neighborhoods with a plane are the union of a constant number of half-ellipses, circles or half-circles and straight line segments. The event points of the sweep over the arrangement are all intersection points of a boundary curve with the triangle $f(\Delta_K)$ and intersection points between two boundary curves that lie inside $f(\Delta_K)$.

The overall time complexity of the second step of the algorithm adds up as follows: For each triangle $\Delta_K$ we need to decide by which connected components of the free space it is completely covered. For one connected component, the arrangement for $\Delta_K$ has size $O(l^2)$ and can be swept in time $O(l^2 \log l)$ where $l$ is the number of cells projecting onto $\Delta_K$. Let $s$ be the number of connected components and $l_i, 1 \leq i \leq s$, the number of cells projecting onto $\Delta_K$ in the $i$th connected component. Because the connected components of the free space are disjoint the sum of the $l_i, 1 \leq i \leq s$, is at most $m$. Thus, for each triangle $\Delta_K \in K$ we can determine in $O(\sum_{i=1}^{s}(m_i^2 \log m_i)) \in O(m^2 \log m)$ time, by which of the connected components of the free space it is covered. We have to apply the same procedure with $K$ and $L$ exchanged. This yields a total runtime of $O(nm^2 \log m + mn^2 \log n)$ for all triangles in both parameter spaces.

Summarizing, we obtain the following algorithm for the decision problem of the weak Fréchet distance.

---
**Algorithm 2**: DecideWeakFrechet$(f, g, \varepsilon)$
---

**Input**: parametrized triangulated surfaces $f, g, \varepsilon > 0$
**Output**: Is $\delta_{wF}(f, g) \leq \varepsilon$?

1 compute the graph $G_\varepsilon$ of the free space diagram $F_\varepsilon(f, g)$
2 **forall** *connected components $A$ of $G_\varepsilon$* **do**
3     **forall** *triangles $\Delta$ in either parameter space* **do**
4         using line sweep decide whether $\Delta$ is completely covered by the component $A$
5     **end**
6 **end**
7 output *true* if a connected component covering all triangles has been found, else output *false*

Lemma 3 and the analysis above yield the following theorem:

**Theorem 3.** *Algorithm 2 decides whether the weak Fréchet distance between two triangulated surfaces with $n$ and $m$ triangles, respectively, is less than a given parameter $\varepsilon$ in $O(nm^2 \log m + mn^2 \log n)$ time.*

## 3.3 Computing the Weak Fréchet distance

The decision algorithm can be extended to a computation algorithm by searching a set of critical values for $\varepsilon$ as is done for the Fréchet distance between polygonal curves [AG95]. We first characterize these critical values and then show how to compute the weak Fréchet distance by searching over the set of critical values.

By Lemma 3 the weak Fréchet distance corresponds to an extensive connected component of the free space. Thus, the weak Fréchet distance equals a value $\varepsilon$ if and only if the free space $F_\varepsilon$ contains such a connected component and it does not for any smaller value of $\varepsilon$. We can characterize the critical values as follows.

1. The combinatorial structure of the connected components changes

2. The projection of a connected component covers a parameter space completely whereas for smaller values of $\varepsilon$ this is not the case although there is no combinatorial change.

We now describe this subdivision of the critical values more precisely:

**Type 1** The combinatorial structure of the free space changes when a boundary cell, i.e., the cell of an edge and a triangle, becomes non-empty. This happens for $\varepsilon$ equal to the distance of the edge and the triangle defining the boundary cell, as stated in equation (2). There are $O(mn)$ such values, each of which can be computed in constant time.

**Type 2** Apart from the combinatorial changes, the projection of a connected component grows monotonously with increasing $\varepsilon$. We distinguish the critical values of this type by the last points in a certain neighborhood covered by the projection, i.e., points $x \in [0,1]^2$ which are in the projection of the free space $F_\varepsilon(f,g)$ and for any $\varepsilon' < \varepsilon$ there is a neighborhood of $x$ which is disjoint to $F_{\varepsilon'}(f,g)$.

A last point covered may be either a vertex (Type 2a), an interior point of an edge (Type 2b) or an interior point of a triangle (Type 2c) of the triangulation of the parameter space, or - if the triangle are not in general position - the last points covered may be a segment (Type 2d).

**Type 2 a** If the last point covered is vertex, then its image (under $f$ or $g$) lies on the boundary of the $\varepsilon$-neighborhood of a triangle. Thus, its image has distance $\varepsilon$ to the triangle. We can compute these critical values by computing for all triangles of the one surface their distance in image space to the vertices of the other surface. Each such value can be computed in constant time and there are $O(mn)$ such values.

**Type 2 b** If the last point covered lies in the interior of an edge of one surface, then the boundaries of $\varepsilon$-neighborhoods of two triangles of the other surface intersect in the image of that point. That is, the point has equal distance to two triangles. There are $O(mn^2 + m^2n)$ such critical values for both parameter spaces.

**Type 2c** If the last point covered lies in the interior of a triangle, then the boundaries of $\varepsilon$-neighborhoods of three triangles intersect in the point. That is, the point has equal distance to three triangles. There are $O(mn^3 + m^3n)$ such critical values for both parameter spaces.

**Type 2d** If the triangles are not in general position it may also happen that the last points covered form a line segment. A detailed analysis shows that this can occur only if two edges, or two triangles, or an edge and a triangle in image space are parallel and $\varepsilon$ is the distance between such a parallel pair. There are $O(m^2 + n^2)$ such values, each of which can be computed in constant time.

Summarizing, we have $O(m^2n + n^2m)$ critical values of types 1, 2a, 2b, and 2d, each of which can be computed in constant time. Furthermore, we have $O(m^3n + mn^3)$ critical values of type 2c.

We could now compute the weak Fréchet distance by computing and sorting the critical values and then doing a binary search over the set of critical values, solving Algorithm 2 in each step. The runtime of this algorithm would be dominated by the sorting of the critical values which takes $O((m^3n+mn^3)\log(mn))$ time.

We can improve the runtime applying parametric search [Meg83] to the $O(m^3n+mn^3)$ critical values of type 2c where the boundaries of three projected cells intersect. For the $O(m^2n+mn^2)$ other critical values we first run a binary search involving the decision algorithm 2 in each step. This part takes $O((m^2n+mn^2)\log(mn))$ time.

More precisely, we have the following algorithm.

---
**Algorithm 3**: ComputeWeakFréchetDist($f,g$)

**Input**: Parametrized triangulated surfaces $f,g$
**Output**: $\delta_{wF}(f,g)$

1. Compute the set $C_1$ of critical values of type 1, 2a, 2b, and 2d
2. Sort the set $C_1$
3. Do a binary search over the set $C_1$ applying Algorithm 2 in each step and branching to larger values if the answer is false and to smaller values if the answer is true
4. Let $c_i$ be the smallest value in $C_1$ for which Algorithm 2 outputs true
5. Do a parametric search over the set $C_2$ of critical values of type 2c in the interval $(c_{i-1}, c_i]$, again applying Algorithm 2 in each step
6. Output the smallest value for which Algorithm 2 outputs true

---

For the parametric search, instead of giving a parallel algorithm for the decision problem, it suffices to use any parallel algorithm whose critical values include the critical values of the decision problem. In our case, we can use a parallel comparison-based sorting algorithm for sorting lexicographically the intersection points of the boundary curves of $\varepsilon$-neighborhoods of triangles of the one surface intersected with a plane containing a triangle of the other surface. Formally, we sort all points of the form

$$\partial\big((\Delta'_1 \oplus B_\varepsilon) \cap E_\Delta\big) \;\cap\; \partial\big((\Delta'_2 \oplus B_\varepsilon) \cap E_\Delta\big),$$

where $\Delta$ is the image of some triangle in one parameter space and $\Delta'_1, \Delta'_2$ images of triangles of the other parameter space. These are $O(m^2n+n^2m)$ many values and their critical values include the critical values of the decision algorithm.

More specifically, we use Cole's variant of parametric search based on sorting [Col87]. His technique yields a runtime of $O((k+T_{\text{dec}})\log k)$ where $T_{\text{dec}}$ is the runtime of the decision algorithm and $k$ is the number of values to be sorted. In our case, $k \in O(m^2n+n^2m)$ and $T_{\text{dec}} \in O(m^2n \log n + n^2m \log m)$. Thus we get a total runtime of $O((m^2n+mn^2)\log^2(mn))$, which proves Theorem 2.